\documentstyle{europhys}

\def\And{{\rm and\ }}

\def\stars{\bigskip\centerline{***}\medskip}

\newif\ifboo \boofalse

\def\Review#1{\boofalse{\it #1},}
\def\Name#1{{\sc #1},}
\def\Vol#1{\ifboo Vol. {\bf #1}\else{\bf #1}\fi}
\def\Year#1{\ifboo #1\else(#1)\fi}

\def\Page#1{\ifboo {\rm p. #1}\else{\rm #1}\fi}

\input epsf
\begin{document}

\euro{48}{5}{561-567}{1999}
\Date{1 December 1999}

\shorttitle{K. Penc \And R. Lacaze: spl(2,1) dynamical supersymmetry etc.}

\title{ spl(2,1) dynamical supersymmetry and suppression of ferromagnetism
in flat band double-exchange models}

\author{
  Karlo Penc\inst{1}\footnote{on leave from  Res. Inst. for Solid 
      State Physics, Budapest, Hungary.}  
  \And Robert Lacaze\inst{1}\inst{2} }

\institute{
     \inst{1} Service de Physique Th\'eorique, CEA-Saclay, 
     91191 Gif-sur-Yvette Cedex, France \\
     \inst{2} ASCI, Bat. 506, Universit\'e Paris Sud, 
     91405 Orsay Cedex, France 
}

\rec{28 April 1999}{in final form 30 September 1999}

\pacs{
 \Pacs{71}{27$+$a}{Strongly correlated electron systems; heavy fermions}
 \Pacs{75}{30Mb}{Valence fluctuation, Kondo lattice, and heavy-fermion 
phenomena}
 \Pacs{03}{65Fd}{Algebraic methods}
}

\maketitle

\begin{abstract}
The low energy spectrum of the ferromagnetic Kondo lattice model on a
$N$-site complete graph extended with on-site repulsion is obtained 
from the underlying spl(2,1) algebra properties in the strong coupling limit. 
The ferromagnetic ground state is realized for 1 and $N$+1 electrons only. 
We identify the large density of states to be responsible for the 
suppression of the ferromagnetic state and argue that a similar 
situation is encountered in the Kagom\'e, 
pyrochlore, and other lattices with flat bands in their one-particle 
density of states.
\end{abstract}

 It is believed that some aspects of the electronic properties of the strongly
correlated transition metal
oxides, like manganites\cite{exp}, 
 can be revealed by considering the 
Kondo-lattice
Hamiltonian ${\cal H}_{\rm KL} = {\cal T} + {\cal H}_{\rm int}$, with
\begin{eqnarray}
  {\cal T} &=& -\sum_{i,j,\alpha} t_{ij} 
    c^{\dagger}_{i\alpha} c^{\phantom{\dagger}}_{j\alpha}, \qquad t_{ij}>0,
  \label{eq:HKLkin}\\
  {\cal H}_{\rm int} &=&
 -\frac{J_H}{2} \sum_{j,\alpha,\beta}  
  c^{\dagger}_{j\alpha} 
   {\bf S}_{cj} \mbox{\boldmath $\sigma$}^{\phantom{\dagger}}_{\alpha\beta} 
   c^{\phantom{\dagger}}_{j\beta}
 + U \sum_{j} n_{j\uparrow} n_{j\downarrow} ,
  \label{eq:HKLint}
\end{eqnarray}
using standard notations. The kinetic part $ {\cal T}$
describes the hopping of conduction electrons on a lattice 
($\alpha=\uparrow,\downarrow$ is the spin index). The interaction part 
${\cal H}_{\rm int}$ includes the intra-atomic ferromagnetic exchange  
between the conduction electrons and localized spin $S_c$ core 
electrons\cite{hund} with $\mbox{\boldmath $\sigma$}_{\alpha\beta}/2$ and ${\bf 
S}_{cj}$ 
as their respective spin operators
($\mbox{\boldmath $\sigma$}$ denotes a vector of Pauli matrices),
 and the on-site Coulomb repulsion $U>0$ between the electrons.
While here we are interested in  $J_H>0$ Hund's coupling, the Kondo-lattice 
Hamiltonian has been extensively studied for 
$J_H<0$, as describing heavy fermion systems\cite{revKL}.

 Because the Hund's coupling is typically larger than the  hopping, 
the energetically unfavored low spin 
states are neglected (that is when electron and core spin are antiparallel),
and we get the quantum double exchange
 \cite{kkubo,mullerhartmann}. 
The next usually used approximation
neglects the quantum spin fluctuations of the $S_c$ core spin described by 
classical variables
(spherical angles $\theta$ and $\phi$). These two approximations lead to the 
double-exchange
model \cite{mullerhartmann,doubleexch,classicalKL} with Hamiltonian 
${\cal H}_{\rm DE}=-\sum \tilde t_{ij} f^\dagger_i f^{\phantom{\dagger}}_j$ , 
which describes charges as noninteracting spinless fermions 
moving in a disordered background of classical spins with effective 
hopping amplitudes
\begin{equation}
 \tilde t_{ij} =  t_{ij} 
  \left[ \cos \frac{\theta_i}{2} \cos \frac{\theta_j}{2} +
\sin \frac{\theta_i}{2} \sin \frac{\theta_j}{2} e^{i(\phi_i-\phi_j)}
             \right].
 \label{eq:doubleExch}
\end{equation}
The charges can freely propagate provided the core spins are aligned and 
therefore ferromagnetism is favored. The main effect of finite $J_H$ is 
to introduce antiferromagnetic
exchange between the core spins, which will hinder the free propagation of the 
charges, resulting
in a competition between ferromagnetic and antiferromagnetic ordering. 
To check the scenario presented above, different numerical methods
are applied and it has been found that on the cubic lattice the 
ferromagnetism is realized for a wide range of electron 
concentration\cite{dagotto}.

While the scenario described above is widely accepted, in this paper we
would like to point out that the structure of the underlying lattice is 
important and we can expect suppression of the ferromagnetic phase 
when the noninteracting one-particle spectrum has nondispersing states 
forming  the so called flat bands, like in Kagom\'e or pyrochlore lattice. 
This might appear surprising as, according to the Stoner criterion, the 
ferromagnetism goes hand in hand with 
large density of states. As it will turn out at the end, the reason
behind this phenomenon is rather simple and will be demonstrated
as follows: We first
derive the low energy effective Hamiltonian in the strong coupling limit. 
Then the underlying spl(2,1) dynamical supersymmetry allows us to solve the
effective model on the $N$-site complete graph referred as ${\cal G}(N)$ and
defined by a infinitely long ranged hopping $t_{ij}=t(1-\delta_{ij})$ 
(on a lattice with $N$ sites, each site has $N-1$ neighbors).  
The algebraic approach turns out to be a very useful tool, 
and in particular allow us to show that all the wave functions 
can be obtained from the noninteracting ones by an extended Gutzwiller 
projection. The breakdown of the standard scenario is already 
present in our toy model: the ground state is singlet apart from the case
of 1 or $N+1$ electrons in the system, and the 
reason lies in the large density of states in the one--particle 
spectrum of ${\cal G}(N)$.
Finally, we will present arguments 
that the same mechanism will destabilize the ferromagnetic phase in the 
lattices with flat bands. 

To derive the effective Hamiltonian we start at $t=0$,
where the different lattice sites decouple.
An empty site has energy $0$; one electron can form with the core spin
either the high ($S_c+1/2$) or low ($S_c-1/2$) spin state, 
with energies $- J_H S_c/2$ and 
$J_H (S_c+1)/2$, respectively; finally two electrons on a site
results in a state with energy $U$ and spin $S_c$. 
This can be summarized by representing the high and low spin states using 
auxiliary fermions $f$ and $d$, respectively, and the 
spins by Schwinger bosons $b_{\alpha}$,
so that ${\bf S}_j = 
\sum_{\alpha\beta}  b^\dagger_{j\alpha} (\mbox{\boldmath 
$\sigma$}_{\alpha\beta}/2) 
b^{\phantom{}}_{j\beta}$ and
\begin{equation}
 c^{\dagger}_{j\uparrow} = 
   \frac{ b^\dagger_{j\uparrow} f^\dagger_j 
        + b^{\phantom{}}_{j\downarrow} d^\dagger_j
   }{\sqrt{2 S_c+1}}, \quad
 c^{\dagger}_{j\downarrow} = 
   \frac{ b^\dagger_{j\downarrow} f^\dagger_{j}
        - b^{\phantom{}}_{j\uparrow} d^\dagger_j}{\sqrt{2 S_c+1}}. 
\end{equation}
The anticommutation relation for electrons requires the
constraint
$
  \sum_\alpha n^b_{j\alpha} - n^f_j + n^d_j = 2 S_c $ 
to be satisfied at each site
and the  interaction part becomes diagonal:
\begin{equation}
  {\cal H}_{\rm int} =  \frac{J_H}{2} 
  \sum_j\left[(S_c+1) n^d_j - S_c n^f_j - n^f_j n^d_j \right]  
   +U \sum_j n^f_j n^d_j .
\end{equation}
Choosing $U=J_H/2$ we can eliminate the  four fermion term $n^f_j n^d_j$.
Now, using standard techniques \cite{oles94},
we can apply a canonical transformation to get the effective 
Hamiltonian which is the expansion in $t/J_H$ around the atomic 
limit and in the lowest energy subspace, where we keep the $f$ 
fermions only, it reads (see also \cite{sarker})
\begin{eqnarray}
  {\cal H}_{\rm eff} 
  &=& 
    - \frac{J_H N_e S_c}{2}
    - \sum_{i,j,\alpha}  \frac{t_{ij}}{2 S_c \! + \! 1}
     f^\dagger_i b^\dagger_{i\alpha} 
     b^{\phantom{\dagger}}_{j\alpha}  f^{\phantom{\dagger}}_j 
  \nonumber \\&&
     -  \sum_{ijk\alpha\beta\mu\nu}
   \frac{t_{ik} t_{kj}}{J_H (2S_c \!+\! 1)^3}
   \left(
      \delta_{\mu\nu}\delta_{\alpha\beta}
       \! - \! \mbox{\boldmath $\sigma$}_{\mu\nu} \mbox{\boldmath 
$\sigma$}_{\alpha\beta}    
    \right) 
   f^\dagger_{i} b^\dagger_{i\alpha} b^\dagger_{k\mu}
   b^{\phantom{\dagger}}_{k\nu}
   b^{\phantom{\dagger}}_{j\beta} f^{\phantom{\dagger}}_j
   +{\cal O}(\frac{t^3}{J_H^2}) . 
 \label{eq:Heff}
\end{eqnarray}
The first order term $\propto t$ in the expansion is equivalent to
the quantum double-exchange Hamiltonian  
\cite{kkubo,mullerhartmann}. It competes with the next order 
term $\propto t^2/J_H$ which is  essentially an 
antiferromagnetic interaction between the spins, and
there is no need to include higher order terms. 
The formula above is equally valid for $S_c=0$,
where it describes the large-$U$ Hubbard model, implying 
that some of the results below apply also to
the $t$-$J$ model.
The procedure can be repeated for general $U$, resulting in a more complicated 
$t^2/J_H$ term \cite{revKL,pertU}.

When specializing to complete graph ${\cal G}(N)$ the effective 
Hamiltonian can be written as
\begin{eqnarray}
 {\cal H}_{\rm eff} 
 &=&  N_e\varepsilon
  -\frac{t}{2S_c+1} \left(1+\frac{2tN}{J_H(2S_c+1)}\right)
  \sum_\alpha F^\dagger_\alpha F^{\phantom{}}_\alpha 
  \nonumber\\&&
   +\frac{2 t^2}{J_H(2S_c+1)^3} 
              \sum_{\alpha,\beta} 
              F^\dagger_\alpha \left( 
                \hat Y \delta_{\alpha\beta}
              + {\bf S} \mbox{\boldmath $\sigma$}_{\alpha\beta} 
              \right) F^{\phantom{}}_\beta 
   +{\cal O}(\frac{t^3}{J_H^2}),
   \label{eq:Heffcg}
\end{eqnarray}
where $\varepsilon=t- J_H S_c/2$, 
and we introduced the operators ${\bf S} = \sum_j {\bf S}_j$, 
$\hat Y = (S_c+1)N-\hat N_e/2$, 
$F^\dagger_\alpha =\sum_{j} b^\dagger_{j\alpha} f^\dagger_j$ and 
$F_\alpha =\sum_{j} b_{j\alpha} f_j$. The nonvanishing anticommutation  
relations between the fermionic operators are
\begin{equation}
  \left\{F^{\phantom{\dagger}}_\alpha, F^{\dagger}_\beta \right\} = 
  \hat Y \delta_{\alpha\beta}
  + {\bf S}\mbox{\boldmath $\sigma$}_{\alpha\beta}. \\
 \label{eq:ar_xed}
\end{equation}
 The bosonic spin operator ${\bf S}$ satisfy the usual su(2) spin algebra and 
 commutes with $\hat Y$. The system closes with
the commutation relations between the fermionic and bosonic operators 
\begin{equation}
 [F^{\phantom{\dagger}}_\alpha,{\bf S}] = 
\frac{1}{2} \sum_\beta \mbox{\boldmath $\sigma$}_{\alpha\beta} 
F^{\phantom{\dagger}}_\beta,
  \quad  
  [F^{\phantom{\dagger}}_\alpha,\hat Y] = -\frac{1}{2} 
F^{\phantom{\dagger}}_\alpha, 
 \label{eq:cr_mixed}
\end{equation}
with their conjugate.
The set of relations (\ref{eq:ar_xed}-\ref{eq:cr_mixed}) define a spl(2,1) 
graded algebra spanned by ${\bf S}$, $\hat Y$, $F^\dagger_\alpha$ and
$F^{\phantom{\dagger}}_\alpha$, $F^\dagger_\alpha$ and 
$F^{\phantom{\dagger}}_\alpha$, 
while ${\bf S}$ and $\hat Y$ span respectively the su(2) and u(1)
subalgebras\cite{rittenberg76}.
Both the electron number and total spin are conserved, but ${\cal H}_{\rm eff}$
does not commute with the $F$'s - spl(2,1) is not a symmetry of eq. 
(\ref{eq:Heffcg}).
Nevertheless ${\cal H}_{\rm eff}$ can be expressed in terms of the Casimir
operators of spl(2,1), su(2) and u(1) -- the so called {\it dynamical} 
supersymmetry.
Once the representations $[Y,S]$ of spl(2,1) are decomposed into multiplets 
$(S,Y)$
of $su(2)\times u(1)$, we know\cite{rittenberg76} on each multiplet the values 
of the different Casimir
operators of spl(2,1), su(2) and u(1) and similarly to the $t$-model 
\cite{kirson,tmodel}
the spl(2,1) representations give the eigenvalues of the effective Hamiltonian.

The generic irreducible representation (irrep) $[Y,S]$ of the spl(2,1) algebra 
is $8S$ 
dimensional and can be labeled by two linearly independent Casimir 
operators\cite{rittenberg76}.
It contains the $(S,Y)$, $(S-1/2,Y+1/2)$, $(S-1/2,Y-1/2)$ and $(S-1,Y)$ 
spin multiplets.
Special cases concern the irrep 
$[Y=S,S]$ which contains only the $(S,Y)$ and $(S-1/2,Y+1/2)$ spin
multiplets (dimension  $4S+1$) and the irrep
$[Y,S=1/2]$ which do not contain the $(S-1,Y)$ multiplet.
Applying operators $F_\alpha$ and $F^\dagger_\alpha$ we 
can walk between the spin multiplets within an irrep. 

The $(S-1/2,Y-1/2)$, $(S,Y)$, $(S-1,Y)$, and $(S-1/2,Y+1/2)$
multiplets of the irrep $[Y,S]$ are eigenstates of 
$\sum_\alpha F^\dagger_\alpha F^{\phantom{\dagger}}_\alpha$
with the eigenvalues
$2Y-1$, $Y+S$, $Y-S$, and 0, respectively.
As the second order of ${\cal H}_{\rm eff}$ is also quadratic in $F$ operators 
and
can be expressed in terms of the Casimir operators,
the spin multiplets $(S,Y)$ in the irrep $[Y',S']$ are eigenstates
of eq.~(\ref{eq:Heffcg}) with (increasing) energies: 
\begin{eqnarray}
 \label{eq:0nes}
  E_{[Y+\frac{1}{2},S+\frac{1}{2}]}(S,Y) \!\!&=&\!\! N_e\varepsilon 
  -2 \left( t + \frac{2t^2}{J_H} \frac{S^{\rm max}+1}{(2S_c+1)^2}\right)
    \left( N-\frac{S^{\rm max}}{2S_c+1} \right)
  \nonumber \\ && 
 +\frac{4t^2}{J_H} \frac{S(S+1)}{(2S_c+1)^3}
  +{\cal O}(\frac{t^3}{J_H^2}),
 \\
  E_{[Y,S]}(S,Y) \!\!&=&\!\! N_e\varepsilon 
        - \left( t + \frac{2t^2}{J_H}\frac{S^{\rm max}-S}{(2S_c+1)^2} \right)
          \left( N - \frac{S^{\rm max}-S}{2S_c+1} \right) +{\cal 
O}(\frac{t^3}{J_H^2}),
 \\
  E_{[Y,S+1]}(S,Y) \!\!&=&\!\! N_e\varepsilon - 
     \left( t + \frac{2t^2}{J_H} \frac{S^{\rm max}\!+\!S\!+\!1}{(2S_c+1)^2} 
\right) 
 \left( N - \frac{S^{\rm max}\!+\!S\!+\!1}{2S_c+1} \right) 
  +{\cal O}(\frac{t^3}{J_H^2}),
 \\
  E_{[Y-\frac{1}{2},S+\frac{1}{2}]}(S,Y) \!\!&=&\!\! N_e\varepsilon ,
 \label{eq:enes}
\end{eqnarray}
where $S^{\rm max}=N S_c + N_e/2$ is the maximum 
total spin. 
The multiplicity $M^{(N)}_{[Y,S]}$ of the irrep $[Y,S]$ in 
an $N$-site system is given by the 
number of times this irrep is contained in $[S_c+1/2,S_c+1/2]^N$ 
(one site irreps reduce to
$[S_c+1/2,S_c+1/2]$). This is determined from the 
branching rule\cite{rittenberg76}, leading to the following recursion 
relation:
\[
 M^{(N+1)}_{[Y,S+1/2]} = \sum_{y=\{0,1/2\}}
   \sum_{s=-S_c-y}^{S-|S-y-S_c|} M^{(N)}_{[Y+y-S_c-1,S-s+1/2]} ,
\]
with $M^{(N)}_{[Y=N(S_c + 1/2),S]}=\delta_{S,N(S_c+1/2)}$ as 
boundary condition.
This formula along with the energy definition in eqs. 
(\ref{eq:0nes})-(\ref{eq:enes})
allows an iterative procedure to build the energy density distribution of the 
model. 

\begin{figure}  
 \epsfxsize=7.5 truecm  
 \centerline{\epsffile{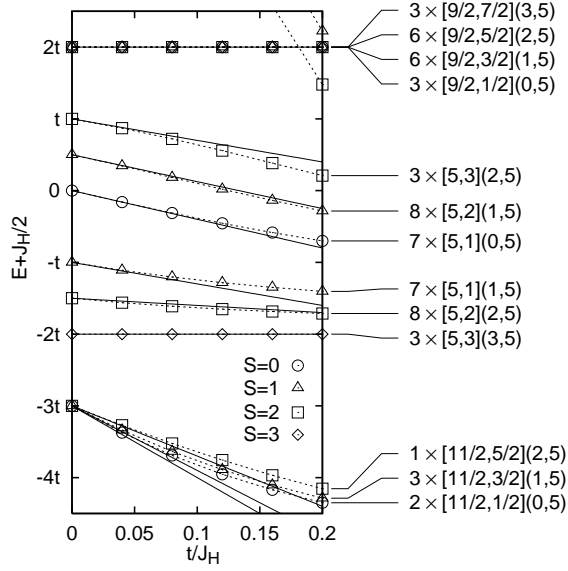}}  
 \caption{The low energy spectrum of the ferromagnetic  
Kondo lattice on a 
4 site complete graph ${\cal G}(4)$ with $S_c=1/2$ and $N_e=2$  
for $U=J_H/2$.  
The solid straight lines show the energy of the effective 
Hamiltonian as given 
by eqs. (\ref{eq:0nes})-(\ref{eq:enes}) to be compared to the dotted lines from
exact diagonalization. The multiplicity and   
the quantum numbers of the irrep the state belong to is also shown.   
At the upper right corner of the plot we can see  
states with low spin $d$ fermions to appear. }  
 \label{fig}  
\end{figure}

In fig.~\ref{fig} the energies given by eqs.(\ref{eq:0nes})-(\ref{eq:enes}) are
compared with those of the Kondo lattice computed by exact diagonalization on a
small ($N=4$) size. At $J_H=\infty$, the observed lowest part of the spectrum is
exactly the one predicted from analysis of spl(2,1) representations.
Up to $t/J_H=0.05$ the effective Hamiltonian energies agree very well with the
exact ones. For larger $t/J_H$ values, higher correction terms need to be
introduced in order to get a quantitative agreement.  As the multiplicities of
the levels shown at the right of fig. 1 do not depend of $t/J_H$ and are those
of the dynamical supersymmetry, these correction can be, in principle, 
calculable.
To obtain this solution it is essential that the effective Hamiltonian 
can be expressed using the operators of the spl(2,1) superalgebra which, for
finite $J_H/t$, is possible for $U = J_H/2$ only (see also \cite{frahm}). 

For $N_e=1$ the $(S,Y)$ spin multiplet of the $[Y+1/2,S+1/2]$ is missing and 
the ground state is the highest spin state in the $[Y,S]$ irrep. 
For $2 \leq N_e\leq N$ the $t^2/J_H$ correction in eq. (\ref{eq:0nes}) makes
the lowest energy state to be the singlet ($S$ integer) or doublet ($S$ 
half--integer),
and the low energy spectrum behaves as $S(S+1)$, like in the 
infinite range, antiferromagnetic Heisenberg model.
The particle-hole transformation can be used for $N_e>N$, and we get that
for $N_e=N+1$ the ground state is again the highest spin state (like 
in the Hubbard model\cite{Tasaki,pieri}).
To summarize, the model is ferromagnetic for $1$ and $N+1$ electrons only.

While the algebraic approach gives the spectrum, 
it does not tell how to get the
wave functions. For the case of the $t$-model it
was shown \cite{simons} by explicit construction that some of the wave
functions can be obtained by Gutzwiller projecting the free fermion ones with 
$\hat P_G = \prod_j(1-n_{j\downarrow} n_{j\uparrow})$. Actually, more is true:
for the quantum double exchange all the large-$J_H$ wave functions can be
obtained by projecting out the $d$ fermions with
$\hat P=\prod_j (1-n_j^d)$ from a suitable set of the wave functions of
the non-interacting Hamiltonian.
To this end, one has to consider the states $|\Phi\rangle$ 
corresponding to the ($S-1/2, Y+1/2$) of the [$Y,S$] irrep. 
They turn out to be exact 
eigenstates of the Kondo lattice Hamiltonian for any value of $J_H$ 
with energy $N_e\varepsilon$ (eq. (\ref{eq:enes})). 
These states satisfy $F_{\sigma}|\Phi\rangle =0$ and thus they are the 
states of the non interacting Hamiltonian which do not contain
zero momentum electron and are invariant with $\hat P$: 
$C_{\sigma}|\Phi\rangle =0$ and $\hat P |\Phi\rangle=|\Phi\rangle$ 
(here $C_{\sigma}=\sum_j c_{j\sigma})$. Starting from these $|\Phi\rangle$ 
and adding zero momentum electrons, we obtain eigenfunctions of the 
$J_H=U=0$ model. Next, using the identity
$\hat P C^\dagger_{\sigma}=F^\dagger_{\sigma}\hat P$,
the projected state remains in the same irrep with wave-functions which is
eigenstate of the large-$J_H$ model. In other words, the extended Gutzwiller
projection (as $\hat P = \hat P_G$ for $S_c=0$) is exact for the model 
on ${\cal G}(N)$.

At this point, it is instructive to study the classical ($S_c \to \infty$) and 
$t^2/J_H=0$ limit of the model. In the language of Schwinger bosons 
 $b^\dagger_{j\uparrow} \approx \sqrt{S_c} \cos(\theta_j/2)$ and
$b^\dagger_{j\downarrow} \approx \sqrt{S_c} \sin(\theta_j/2) e^{i \phi_j}$ 
\cite{shibaprivate}.
This immediately leads to the hopping amplitudes (\ref{eq:doubleExch}) 
of the double exchange model with the one-particle Hamiltonian 
$ {\cal H}_{\rm 1p}=\varepsilon 
             - t (|c \rangle \langle c| + |s \rangle \langle s|)$  where
$|c\rangle = \sum_j \cos\frac{\theta_j}{2} f^\dagger_j |0\rangle$ and
$|s\rangle = \sum_j \sin\frac{\theta_j}{2} e^{i \phi_j} f^\dagger_j|0\rangle$.
If we choose the $z$-axis to point in the direction of the total core spin, the
$|c\rangle$ and $|s\rangle$ are orthogonal and eigenvectors of
${\cal H}_{\rm 1p}$ with energies 
\begin{equation}
 \varepsilon_c=\varepsilon-\frac{t}{2}(N+\frac{S}{S_c})\ , \ 
  \varepsilon_s=\varepsilon-\frac{t}{2}(N-\frac{S}{S_c}) .
\label{eq:eces}
\end{equation}
Furthermore there are $N-2$ states with energy $\varepsilon$. 
While in the lowest energy state the energy is linearly decreasing with $S$, 
and the fermions can freely propagate when the core spins are parallel,
for $|s\rangle$ the tendency is reversed: energy is higher for larger $S$. 
 Filling the one-particle levels, the spectrum for 
$2 \leq N_e\leq N$ electron is $
-N t + N_e \varepsilon $,   
$\varepsilon_c +(N_e-1) \varepsilon$,  
$\varepsilon_s +(N_e-1) \varepsilon$, and $N_e \varepsilon$.
These energies are equal ${\cal O}(1/S_c)$ to the energies 
(\ref{eq:0nes})-(\ref{eq:enes}),
respectively, and thus the
correspondence between the semiclassical and quantum spectra for this
  model is established. 
The absence of the true ferromagnetic state is due to the 
cancellation of the contributions linear in $S$ for $N_e \geq 2$.
 Again, the projection operator establishes a relationship between 
$J_H=0$ and $J_H \to \infty$ states: 
$|c\rangle = \hat P C_\uparrow |0\rangle$ and 
$|s\rangle = \hat P C_\downarrow |0\rangle$, and the role of the core 
spins is to act like an effective magnetic field which Zeeman splits the 
energy of $C^\dagger_\sigma |0\rangle$ states.

 Now, once we have seen that the ferromagnetism is suppressed in the model,
we can search the reason of such a phenomenon. One of the peculiarities
of our model is the large degeneracy of the one-particle spectrum when 
all the spin are aligned. Then $S$ is maximum and we have one state at
$E=\varepsilon-tN$ and $N-1$ at $E=\varepsilon$. 
Turning over one of the 
core spins the $S/S_c$ ratio is reduced by 2, increasing the lowest energy 
$\varepsilon_c$ by $t$, decreasing the energy of one of the levels in the 
degenerate manifold with the same amount $t$ (actually 
$\varepsilon_s$ in eq.~\ref{eq:eces}) and
leaving the other $N-2$ states of the degenerate manifold with the same energy 
$\varepsilon$. 
In other words, we can gain 
antiferromagnetic energy without loosing total kinetic energy. For this to 
happen, 
the existence of the macroscopically large number of nondispersing 
states is necessary. 
This feature is also present in models defined on lattices of  
corner shared complete graphs 
(known as line graps, see {\it e.g.} \cite{Mielke}). 
Let us call ${\cal G}_{\bf r}(M)$ the $M$-site 
complete graphs centered around the point ${\bf r}$ and first 
concentrate on the family where in $D$ dimensions ${\cal G}_{\bf r}(D+1)$ 
are sharing their 
corners, and (anti)periodic boundary condition are assumed. 
In $D=2$ it is represented by the Kagom\'e  and in $D=3$ by the 
pyrochlore lattice.
In momentum space they have two dispersing bands and just above them 
$D-1$ flat bands. We start to fill the flat bands above 
the electron density $n_C=2/(D+1)$, and we can expect the mechanism we 
outlined above to act when we flip some of the core spins. 
For $D \geq 3$ we can go even 
further following Ref.~\cite{brandt}: the Hamiltonian for $t^2/J_H=0$ can be 
written 
as ${\cal H} = {\cal E}(N_e) + {\cal R}$, where the number ${\cal E}(N_e)$ is
$ -J_H S_c N_e/2 -4t(N-N_e) (1+S_c)/(1+2S_c)$ while
 ${\cal R}=
  t/(2S_c+1) \sum_{\alpha,\bf r} F^{\phantom{\dagger}}_{\alpha,{\bf r}}  
  F^\dagger_{\alpha,{\bf r}}$ is a positive semidefinite operator. The 
summation is over the centers of the ${\cal G}_{\bf r}$ constituting the 
lattice
with corresponding fermionic operator $F_{\alpha,{\bf r}}$. Choosing an
initial state $|\Phi\rangle$ such that
$F_{\alpha,{\bf r}}  |\Phi\rangle =0$ for any ${\bf r}$, the 
$|\Psi\rangle = \prod_{\bf r} F^\dagger_{\uparrow,{\bf r}} 
F^\dagger_{\downarrow,{\bf r}} |\Phi\rangle$   
is an exact ground state of the model with 
energy ${\cal E}(N_e)$, as ${\cal R}|\Psi\rangle=0$.
The number of good $|\Phi\rangle$ states is large 
(e.g. $(2S_c+1)^N$ if it does not contain electrons) and  
$|\Psi\rangle$ will inherit its degeneracy and spin.
Our wave function $|\Psi\rangle$ can be visualized as the product of suitable 
chosen ground states (eq.~\ref{eq:0nes}) of each graph ${\cal G}_{\bf r}$. 
 For finite $t^2/J_H$ the 
antiferromagnetic term will split the  
$|\Psi\rangle$ manifold and a high spin state will certainly not be the
ground state for electron densities larger than
$n_Q=4/(D+1)$ (we added two electrons for each ${\cal G}_{\bf r}$ by 
constructing $|\Psi\rangle$).  
The construction can be repeated for 
other line graphs as well. For example when ${\cal G}_{\bf r}(2D)$ form
a $D$-dimensional hypercubic lattice\cite{Mielke}, we have one dispersing 
bands and $D-1$ flat bands and, consequently, the critical densities are 
lower, $n_C=1/D$ and  $n_Q=2/D$. It should be noted here that in the 
cubic lattice the ferromagnetism is suppressed near zero and half fillings 
due to lack of available carriers\cite{doubleexch,dagotto}, and it is not the 
case here, where charges 
are available, however in dispersionless states.

From the arguments above  
the following speculative phase diagram emerges. (i) For densities
below $n_C$ the usual double exchange mechanism will stabilize the 
ferromagnetic phase in a large parameter range. 
(ii) For densities between $n_C$ and $n_Q$ preliminary studies indicate a kind 
of 
ferrimagnetic state. (iii) Finally, if $n>n_Q$ the physics is governed by the 
antiferromagnetic term $\propto t^2/J_H$ in the effective Hamiltonian 
with a possible spin liquid ground state.

  Finally, let us compare our result with the Stoner model. The Stoner model
is a molecular field model of itinerant electrons, where the magnetization is 
gained at the expense of the kinetic energy, thus large density of states is 
favored. While certainly we cannot apply a mean field model to strongly 
correlated system, it turns out that in the Hubbard model singularities
in the density of states can help the ferromagnetism to develop in some 
particular cases \cite{vollhardt}.
Therefore one is tempted to generalize the conclusions drawn from the Stoner 
model. Here we have shown that the flat bands (at least if they are above 
the broad dispersing bands) inhibits ferromagnetism in the 
Kondo lattice. If we look more carefully, the fact that the driving force 
for finite magnetization is the kinetic energy gained by spin alignment, we 
do not expect from the beginning the Stoner criterion to apply.

  To summarize, we have shown that the strong coupling limit of the 
Kondo lattice with infinite long range hoppings can be solved exacly 
using the underlying dynamical supersymmetry. We learned that on this 
particular lattice (i) an extended Gutwiller projection becomes exact, (ii)
the ferromagnetic ground state is not favored. Finally, extending our
result to more general lattices we arrived at an interesting conjecture 
that large density of states is against ferromagnetism in the double 
exchange model.

\stars

 We would like to acknowledge useful discussions with P. Fazekas,
F. Mila, E. M\"uller-Hartmann and H. Shiba.

\end{document}